\documentclass{article}

\usepackage{graphicx}				
\usepackage{amsmath}				
\usepackage{amsfonts}
\usepackage{url}
\usepackage[colorlinks=true,linkcolor=blue,anchorcolor=black,citecolor=blue,filecolor=black,menucolor=black,urlcolor=blue,breaklinks=true,pdfhighlight=/P,pdfmenubar=true,pdftoolbar=true,pdfpagelabels=true,pdfstartpage=1,pdfstartview=FitV,pdftitle={Crowd Research at School: Crossing Flows},pdfsubject={Pedestrian, Crowd, Mass, Simulation, Dynamics, VISWALK, VISSIM},pdfauthor={Kretz},pdfcreator={tobias Kretz},pdfproducer={t kretz},pdfkeywords={Pedestrian, Crowd, Mass, Simulation, Dynamics, VISWALK}]{hyperref}	
\usepackage[numbers,sort&compress]{natbib}	
\usepackage{hypernat}				
\usepackage[american]{babel}		
\usepackage{simplemargins}
\setallmargins{1in}
\usepackage[american]{babel}		

\begin{document}

\title{Crowd Research at School: Crossing Flows}
\author{Johanna Bamberger$^1$ \and Anna-Lena Ge{\ss}ler$^1$ \and Peter Heitzelmann$^1$ \and Sara Korn$^1$ \and Rene Kahlmeyer$^1$ \and Xue Hao Lu$^2$ \and  Qi Hao Sang$^2$ \and  Zhi Jie Wang$^2$ \and  Guan Zong Yuan$^2$ \and  Michael Gau{\ss}$^3$ \and Tobias Kretz$^4$\\ \\
$^1$: Bismarckgymnasium, D-76133 Karlsruhe, Germany \\
$^2$: Jing Ye High School, Shanghai, China \\
$^3$: Fernstudienzentrum -- Karlsruhe Institute of Technology (KIT), D-76133 Karlsruhe, Germany\\
$^4$: PTV Group, D-76131 Karlsruhe, Germany \texttt{Tobias.Kretz@ptvgroup.com}}

\maketitle

\abstract{It has become widely known that when two flows of pedestrians cross stripes emerge spontaneously by which the pedestrians of the two walking directions manage to pass each other in an orderly manner. In this work, we report about the results of an experiment on crossing flows which has been carried out at a German school. These results include that previously reported high flow volumes on the crossing area can be confirmed. The empirical results are furthermore compared to the results of a simulation model which succesfully could be calibrated to catch the specific properties of the population of participants.}

\section{Introduction}

\subsection{Background: the Science Exchange Program Shanghai - Karlsruhe}
The Science Exchange Program Shanghai - Karlsruhe is a student exchange program for students at an age of 16 to 17 years and come from various high schools from Karlsruhe as well as Shanghai's Jing Ye high school. It is biannually jointly organized  by the Sch{\"u}lerakademie Karlsruhe and the Karlsruhe Institute of Technology (KIT). In two weeks of which one is held in Karlsruhe and one in Shanghai teams of four to eight pupils (half of them Chinese and half German) work on particular projects which are supervised by University groups, companies or research centers. In 2012 one group decided to analyze crossing pedestrian flows as their project for their week in Karlsruhe. The main part of the project was to carry out an experiment. Some simulations were done, but for this contribution additional simulations were carried out after the end of the actual student exchange program project.

\subsection{History of the Research on Crossing Pedestrian Flows}
In 1977 Naka \cite{Naka1977} found that in crossing pedestrian flows stripes form which are orthogonal to the sum of the velocity vectors of the two main walking directions. This was confirmed in 1988 by Ando, Ota, and Oki \cite{ando1988forecasting}. However, while in the paper by Naka the stripes are sketched\footnote{One can assume that most readers outside Japan will only have understood the sketches in both papers, as both papers are written in Japanese.} more as extended bubbles, with a rather organic appearance, in the later paper the sketch is drawn with straight lines, strictly and clearly separating pedestrians of the two walking directions. This sketch was used in English language papers for example by Hughes \cite{hughes2002continuum} or Helbing et al. \cite{helbing2005self}. In this way the idea of strict and straight stripes spread and was eventually even popularized (e.g. Fisher \cite{fisher2009perfect}). Seeing literature evolve in this way the question emerges if the blurry, bubble-shaped stripes describe reality best and the straight stripes are maybe an urban myth resulting from a kind of Chinese Whispers Effect (here rather ``Japanese Whispers Effect'') or if stripes in crossing flows can actually be seen as clearly as the sketches in more recent popular publications suggest. Making another attempt to decide this question was the purpose of the experiment carried out as part of this project. Furthermore we wanted to test Naka's claim that ``The total flow rate of two flows in the crossing can reach nearly the maximum of the flow rate of one direction flow''.

Further notable works on crossing pedestrian flows include: an empirical study by Sano and Tatebe \cite{sano2000difficulty}; extensive experiments on crossing flows of various angles by Wong et al. and calibrated a simulation model with the data which then gave decreasing flows with increasing intersection angle (i.e. bi-directional flow is more reduced than orthogonally crossing flow) \cite{wong2010bidirectional}; a work on potential benefit of crowd control by Yamamoto and Okada \cite{yamamoto2011continuum} where both, a macroscopic continuum model as well as a microscopic simulation model, yielded rather clear and straight stripes; Dias et al. found oscillating flows when two groups of ants meet in a crossing flow situation \cite{dias2012intersecting}; in a contribution with a theoretical focus B{\"a}rwollf et al. also present some empirical data which gives the impression that rather blurry stripes or oscillations were observed \cite{baerwolff2014}. We also point out  the contributions of Cividini et al. \cite{cividini2014} and Duives et al \cite{Duives2014} in this proceedings book.

\section{Experiment}
\subsection{Experimental Setup}
\begin{figure}[htbp]%
\center
\includegraphics[width=0.3\columnwidth]{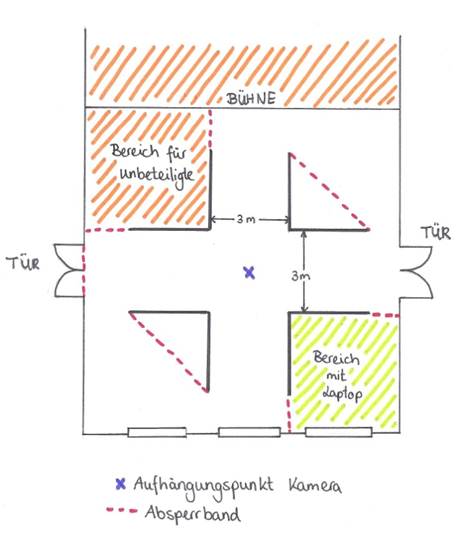}%
\includegraphics[width=0.3\columnwidth]{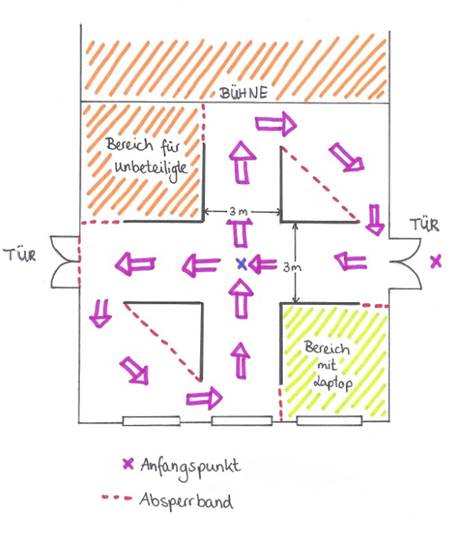}%
\includegraphics[width=0.3\columnwidth]{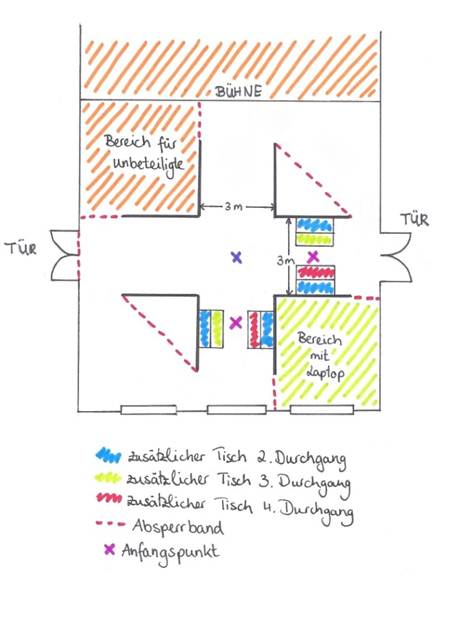}%
\caption{Sketch of geometrical setup.}%
\label{fig:setup}%
\end{figure}

The participants were a group of 80 pupils most of them at an age of 13 or 14 years. The walking geometry as shown in figure \ref{fig:setup} was about-8-shaped with a crossing area of 3x3 sq m. In the first run the access corridors had the same width (3 m) as the crossing area, in the second run two desks were added reducing the width to a total of 2.0 m, in the third run one more desk was added, resulting in a total access width of 1.5 m and in the fourth run four desks left an access width of 1.0 m. The camera was placed in a height of approximately 3.5 m above the center point of the crossing area. This was just sufficient to cover the entire crossing area.

\subsection{Results of the Experiment}
\begin{figure}[htbp]%
\center
\includegraphics[width=0.612\columnwidth]{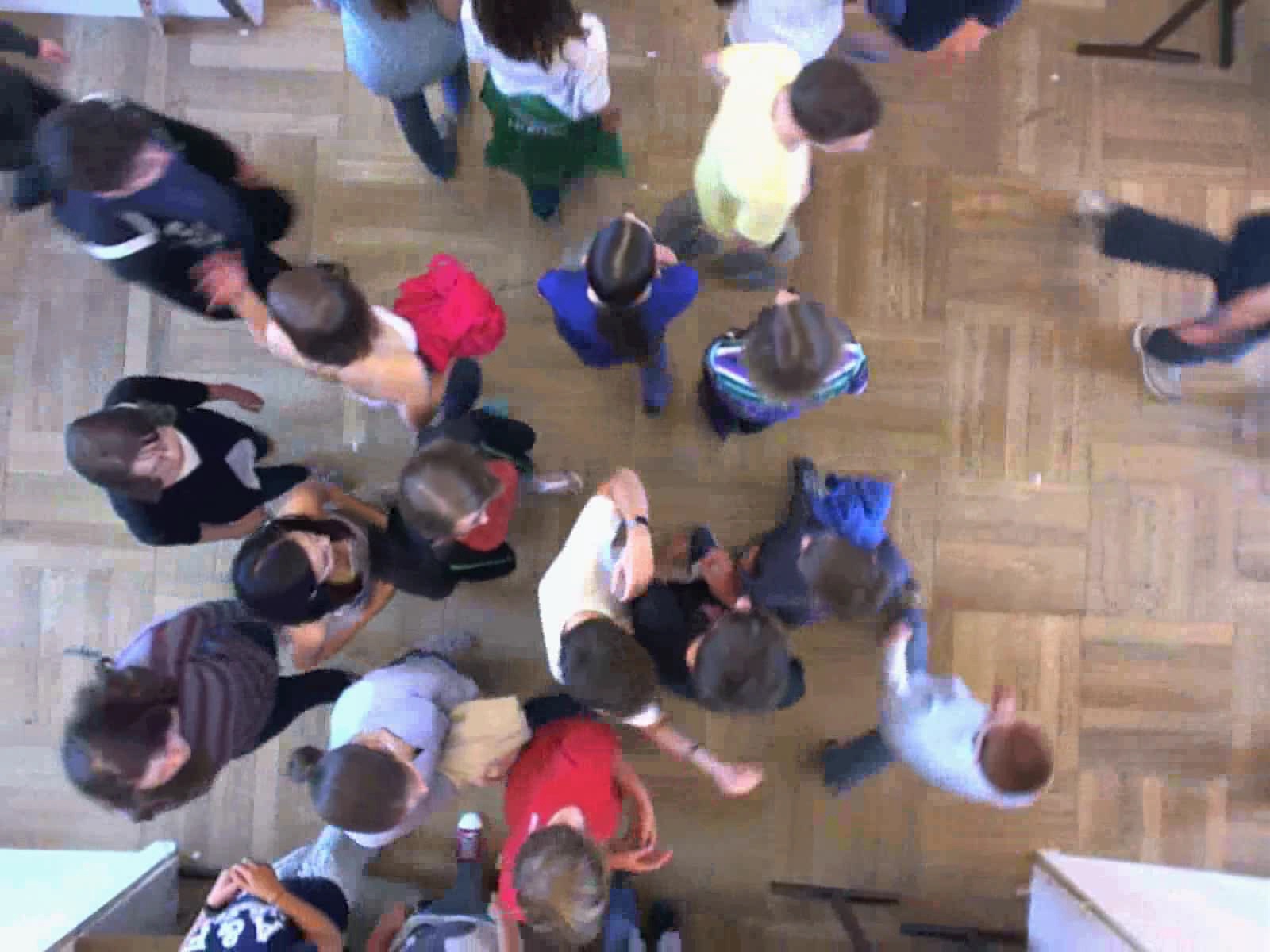}%
\caption{Still image from the third run with three desks restricting flow into the crossing area.}%
\label{fig:experiment}%
\end{figure}

\begin{figure}[htbp]%
\center
\includegraphics[width=0.612\columnwidth]{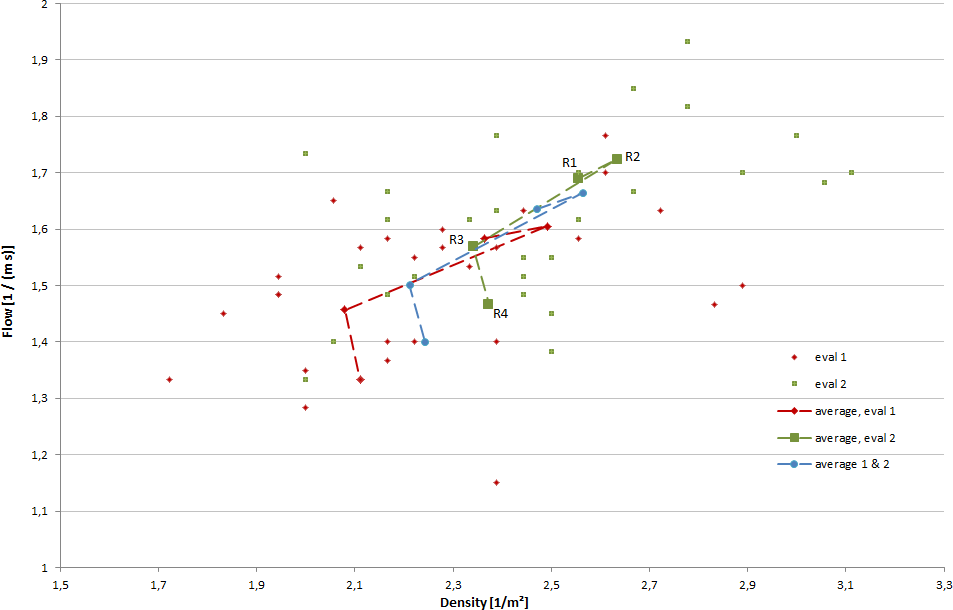}%
\caption{Results of manual counts of two persons (red diamond and green squares) as well as their averages for each run (R1 - R4). Note that the lower left corner of the diagram is not the offspring. The blue circles mark the average of both counts. It is remarkable how stable the bias between the two counting persons is. One may conjecture that it is more difficult to estimate if a person is still on the crossing area or right next to it than it appears and that each counting person has had different lines in mind to judge this.}%
\label{fig:results}%
\end{figure}

Figure \ref{fig:experiment} shows a still image from the third run of the experiment. The evaluation was done manually. For this the video footage was partitioned into segments of 20 seconds.  In these 20 seconds the number of participants leaving the crossing area on the two outbound cross sections has been counted. This number divided by 3 meters and 20 seconds is the specific flow of that particular segment. For density we have counted the number of pedestrians on the crossing area at times t=5 s and t=15 s of the particular segment. The results are displayed in figure \ref{fig:results}. Segments where the participants obviously were doing other things than walking were not taken into account.

The results are not easy to be interpreted. With decreasing access width at first density and flow increase, then they decrease and finally the flow is decreased while the density is increased. This could be related to a learning or a tiring effect or both. However, indisputably the specific flow is high, as Naka stated ``nearly the maximum of the flow rate of one direction flow''. Concerning the organization of the flow we report that to us it appeared that there were rather bubbles moving, almost oscillations of the right of way. Sometimes one could realize rather clear stripes, but these were stable only for a few seconds -- about as long as it takes for a person to cross the crossing area.

\section{Simulation}
It is a natural next question if the empirical results can be reproduced with a simulation model of pedestrian dynamics. We have tried to do so using PTV Viswalk \cite{Kretz2008b,Boenisch2009,sahaleh2012scenario,Kretz2013e} and in a first attempt achieved slightly too small densities and by far a too small flow. This is, however, not a surprise, as we could not expect to have the simulation reproduce results with this special population of participants applying nothing but the default parameters. The first, obvious modification was to make the simulated pedestrians smaller than by default as the participants were all children or young youths. When we applied the changes as shown in table \ref{tab} the results matched much better as can be seen in figure \ref{fig:simresults}.

\begin{table}
\center
\begin{tabular}{l|rr}
Parameter & Default & Calibrated \\ \hline
$v_0^{min}$       & 0.72 m/s    & 1.25 m/s    \\
$v_0^{max}$       & 1.61 m/s    & 1.53 m/s    \\
body radius (min) & 16 cm       & 11 cm       \\
body radius (max) & 23 cm       & 19 cm       \\
tau               & 0.40 s      & 0.37 s      \\
$A_{soc,iso}$     & 1.6 m/s$^2$ & 2.2 m/s$^2$ \\
$A_{soc,mean}$    & 0.4 m/s$^2$ & 0.2 m/s$^2$ \\
$\Delta$t (aka VD)& 3 s         & 2 s         \\
\end{tabular}
\caption{Parameter modifications for calibration. Within the bounds of desired speed ($v_0$) and body radius the values were equally distributed in either case.}
\label{tab}
\end{table}

We have noted that it is crucial for the movement pattern how wide the distribution of desired speeds is. If everyone has (nearly) identical desired speed the emerging stripes are clear, straight, and stable. The wider the distribution is the more the stability is reduced and a deadlock consequently occurs at lower density and flow. This is -- by the way -- also the case for bi-directional flow for which this relation has recently been found as well in an empirical investigation \cite{moussaid2012traffic}. 

It is interesting to note that the values of the the $A$ parameters had to be changed into different directions. $A_{soc,iso}$ sets the base impact of the variant of the Social Force Model as introduced in \cite{helbing2000simulating} while $A_{soc,mean}$ determines the strength of the elliptical specification II as introduced in \cite{johansson2007specification}. This necessity for divergent modification shows how different the two variants actually are. While most parameter changes had an effect on either the density on the crossing area or the flow or both, interestingly simulation results were nearly insensitive to the choice if pedestrians followed a static or a dynamic potential \cite{Kretz2011e} over the crossing area -- a difference which has been found to have a heavy impact on the course of simulations of other situations \cite{Kretz2011f,Kretz2012e,Kretz2013d,Kretz2014a}

Note that the calibration process has been carried out with regard to the empirical data of the first evaluating person. It could have been done as well according to the average or the result of evaluating person 2, but considering the large scattering of empirical data it does not make sense to attempt to achieve a very precise agreement of simulated and empirical data as obviously the empirical data yield considerable variation and potential errors. We therefore see it as sufficient agreement when the average of simulation data is within the cloud of empirical data points.

Excerpts from a simulation run with clear stripes (narrow speed distribution), excerpts from the experiment, and excerpts from the calibrated simulation can be seen in a video at this URL: \url{http://youtu.be/vdjbIYpa1XE}

\begin{figure}[htbp]%
\center
\includegraphics[width=0.612\columnwidth]{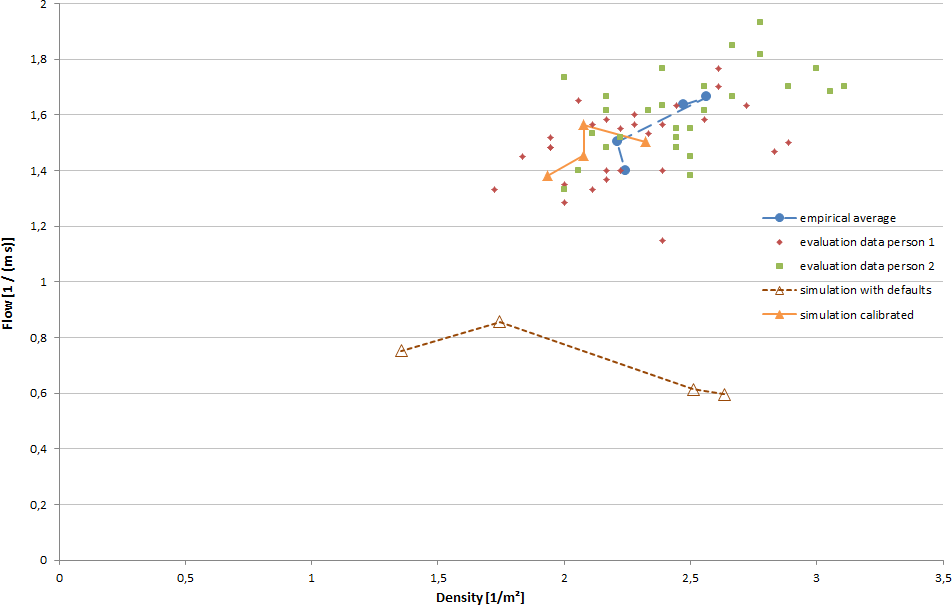}%
\caption{Simulation and empirical results in comparison.}%
\label{fig:simresults}%
\end{figure}

\section{Conclusions}
As a summary we conclude that in our experiment on crossing pedestrian flows the formerly reported high flows and densities were confirmed. We did not see, however, the formation of clear and stable stripes. The movement pattern most of the time was rather bubble-shaped towards oscillating flow directions. When there were stripes they disappeared again soon. This does not mean that there are no situations with more stable stripes. It could be that stripes emerge more clearly when the crossing area is larger. One can also argue that our population of participants was rather individualistic and rather not willing to adapt walking speeds mutually (remembering that simulations had shown that narrow distributions of desired speeds lead to a more pronounced stripe formation). Furthermore the corridors leading to the crossing area were rather short such that there was no time for the pedestrians to pre-sort themselves. Finally the demand (the inflow) on the crossing area was relatively high in all four runs. Only in the last run the access width was below half of the corridor width. This means that in the first three runs in sum demand met or was above the capacity of the crossing area and probably this was the case even in the last run with an access width of 2 x 1 m. Maybe stripes emerge rather at demands clearly below capacity.

In other words as hints for anyone who might want to do a similar experiment we note that first: it would have been interesting to do the experiment with an even smaller inflow opening width, second: to potentially allow the formation of stripes a larger crossing area was required, third: it would have been helpful for evaluation to record a larger environment of the crossing area, fourth: 80 participants was just enough or rather few for this geometry, fifth: the inflow regulating obstacles (tables) should better have been placed further upstream to allow the stream to fully expand until it reaches the crossing area.

The conclusion concerning the simulation is that it is possible to reproduce the empirically observed behavior to that degree of detail to which we analyzed the empirical data plus a visual comparison of movement patterns. The parameters of the simulation had to be adapted for this. The default parameters led to -- compared to our experiment and our population of participants -- clearly more conservative results.

As a last conclusion we would think that the science of pedestrians (and vehicles) can contribute in many and interesting ways to education at school. Pupils and students at secondary schools are involved with the topic in various ways in their everyday lives and the systems as such are well suited to introduce with them a number of concepts of mathematics and computer science.

\section*{Acknowledgements}
We thank Yusuke Okahira from Kozo Keikaku Engineering (KKE) for help with some of the Japanese papers.


\bibliographystyle{utphys2011}
\bibliography{tgf13-crossing-flow}

\end{document}